\documentclass[12pt]{iopart}
\usepackage{graphicx}
\usepackage{cite}
\begin{document}
\submitto{J.Phys.Cond.Mat.}\title[Dynamics and depinning of the triple contact line]{Dynamics and depinning of
the triple contact line in the presence of periodic surface defects}
\author{Vadim S. Nikolayev}
\address{ESEME, Service des Basses Temp\'eratures, DRFMC/DSM, CEA-Grenoble, France\footnote{Mailing
address: CEA-ESEME, PMMH, ESPCI, 10, rue Vauquelin, 75231 Paris Cedex 5, France}}
\ead{vnikolayev@cea.fr}
\date\today

\begin{abstract} We propose an equation that describes the shape of the driven contact line in dynamics
in presence of arbitrary (possibly random) distribution of the surface defects. It is shown that the
\emph{triple} contact line depinning differs from the depinning of interfaces separating \emph{two} phases; the
equations describing these phenomena have an essential difference. The force-velocity dependence is considered
for a periodical defect pattern. It appears to be strongly non-linear both near the depinning threshold and for
the large contact line speeds. These nonlinearity is comparable to experimental results on the contact line
depinning from random defects.
\end{abstract}

\section{Introduction}

Motion of interphase boundaries in a random environment remains an open problem of general
interest. Much attention has been paid to the depinning transition in the systems where collective
pinning creates non-trivial critical behaviour of the interface separating two different phases:
fluid invasion in porous media, magnetic domain wall motion, flux vortex motion in type II
superconductors, charge density wave conduction, dynamics of cracks, solid friction
\cite{Stanley,Fisher}. The theory of the depinning transition is based on the analysis of the
following equation for the interface position $h$:
\begin{equation}\label{depin}
\frac{\partial h}{\partial t}=F+\eta(h)+G[h],
\end{equation}
where $F$ is the externally imposed force, $\eta$ is the noise due to the randomness of the media,
$t$ is time, and $G[\cdot]$ is some operator. When $F$ is close to the depinning threshold $F_c$
(where the interface begins to move), this approach generally results in the power law for the
average interface velocity $v$
\begin{equation}
v\sim(F-F_c)^\beta,  \label{vc}
\end{equation}
where the exponent $\beta$ is universal. The origin of this dependence lies in the peculiar
interface dynamics near the pinning threshold namely a random succession of avalanches of depinning
events. When $F\gg F_c$, a conventional mobility law
\begin{equation}
v\sim F  \label{vf}
\end{equation}
becomes valid.

Depinning of the triple gas-liquid-solid contact line on a solid surface with defects is another
example of the depinning transition. The outlined above general approach to the interface depinning
phenomena is frequently applied to the contact line depinning \cite{EK,Wong,Krauth,LeDoussal}.
However, the discrepancy between the theory and the experimental data on contact line motion is
notable. First, $\beta<1$ according to the theoretical studies (see \cite{EK,LeDoussal}), while
$\beta\geq 1$ was found experimentally \cite{Wong,Moul}. Second, the linear regime (\ref{vf}) was
never obtained \cite{Moul}.

In this paper we propose a framework suitable to explain these results.

\section{Modeling of the contact line motion}

The major problem in this field is related to the failure of the conventional hydrodynamic approach
based on the "no-slip" boundary condition (zero liquid velocity) at the solid surface in the
vicinity of the contact line. Such an approach \cite{Huh} would result in a mathematical
singularity at the contact line: the diverging viscous dissipation. In reality the dissipation in
the vicinity of the contact line is large but finite \cite{DeGennes}. The mechanism of the
singularity removal for the partial wetting case is still under debate. Multiple singularity
removal mechanisms were proposed \cite{Rame,YP}. Most of these models (those which are not limited
to the small values of the dynamic contact angle $\theta$) result in the following expression for
the contact line velocity $v_n$
\begin{equation}
v_n = \frac{\sigma}{\xi}(\cos\theta_{eq}-\cos\theta),  \label{cos1}
\end{equation}
where $\theta_{eq}$ is the equilibrium (Young) value of the contact angle, $\sigma$ is the surface
tension, and $\xi$ is a mechanism-dependent coefficient that has the same dimension as the shear
viscosity $\mu$.

Since the contact line is not straight due to the presence of the defects, the theoretical analysis of pinning
requires the 3D modelling. Such a modelling can be extremely difficult when using the hydrodynamic contact line
motion models where the flow pattern depends on the contact line curvature. Recent papers \cite{PRE02,EuLet03}
have introduced a simpler approach. It is valid if the most dissipation that occurs in the fluid with the moving
contact line takes place in the near contact line region. The latter is defined as a fluid "thread" adjacent to
the contact line. Its diameter is assumed to be much smaller than the radius of curvature of the fluid surface.
In other words, the viscous dissipation in the bulk of the fluid is neglected with respect to the dissipation
that occurs close to the contact line. This assumption is verified by numerous experiments (see e.g.
\cite{JFM,Lang04}).

A single constant dissipation coefficient $\xi$ is introduced to account for the anomalous
dissipation in the vicinity of the contact line without detailing its origin. By further assuming
that this dissipation is the same both for advancing and receding contact line motion, the energy
dissipation rate can be written in the lowest order in $v_n$ as \cite{DeCon}
\begin{equation}\label{diss}
T=\int\frac{\xi \, v_n^2}{2}\;{\rm d}l,
\end{equation}
where the integration is performed along the contact line. The equation for the contact line motion
can be obtained from the force balance between the induced and friction forces \cite{DeGennes}
\begin{equation}\label{var}
  -{\delta U\over \delta h}={\delta T\over \delta \dot h},
\end{equation}
where $\delta\ldots\over\delta\ldots$ means functional derivative, dot means the time derivative,
and $h$ defines the contact line position. The potential energy $U$ of the system needs to be
calculated assuming that each time moment the fluid surface takes its equilibrium shape. It was
shown recently \cite{Iliev} that such a quasistatic scheme leads to Eq. (\ref{cos1}) for an
arbitrary contact line geometry. The dynamic approach (where the fluid surface shape is determined
from hydrodynamics) is considered elsewhere \cite{Sergey}. It appears to lead to the same
Eq.~(\ref{cos1}).

The quasistatic approach is of course not new. It was applied by many researchers, in particular by
Golestanian and Rapha\"el in the approximation of the small contact angles \cite{Raph}. The
advantage of our approach is the
 account of the gravity or/and fluid volume conservation which allow to obtain
the contact line shapes rigourously. In particular, for the Wilhelmy geometry considered below, we
take into account the gravity influence which permits \cite{EuLet03} to avoid divergences
\cite{DeGennes} and thus obtain the contact line profile. When the gravity is irrelevant (as for
small drops), the fluid volume conservation plays this stabilising role \cite{PRE02,Iliev}.

\section{Contact line equation for the forced motion}\label{sec3}

\begin{figure}
  \begin{center}
  \includegraphics[height=4cm]{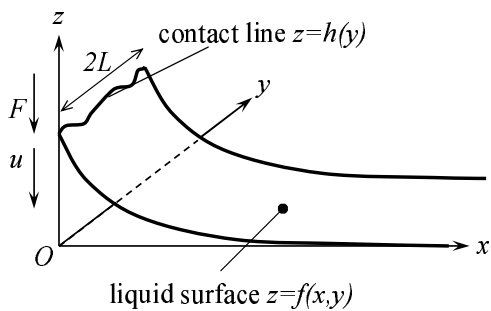}
  \end{center}
\caption{Reference system to describe the Wilhelmy balance experiment. The Wilhelmy plate is
positioned in $yOz$ plane. The positive directions for $u$ and $F$ are shown too.} \label{Wilh}
\end{figure}

The equation for the spontaneous motion has been derived in \cite{EuLet03}. In this paper we deal
with the Wilhelmy geometry (Fig.~\ref{Wilh}), where the vertical plate with surface defects can be
moved up and down with a constant velocity $u$ ($u>0$ for the advancing contact line is assumed).
The average value of the force $F$ exerted on this plate due to the presence of the moving contact
line can be measured with a high precision \cite{Moul}. The liquid-gas interface is assumed to be
described by the function $z=f(x,y,t)$ where $t$ is time and
\begin{equation}\label{cond} |\nabla f|\ll 1
\end{equation} is assumed. The position of the contact
line is then given by its height $h=h(y,t)$ such that $h(y)=f(x=0,y)$. From now
on, we omit the argument $t$.

Under the assumption (\ref{cond}), the minimization of the potential energy $U$ of the liquid  with
respect to $f$ results \cite{EuLet03} in the following expression:
\begin{equation}
f=\frac{1}{2L}\sum_{n=-\infty}^\infty\exp\left(-x\sqrt{l_c^{-2}+{\pi^2 n^2}/{L^2}}\right)
\int_{-L}^L\textrm{d}y'h(y')\cos{\pi n(y-y')\over L},\label{f}
\end{equation}
where $l_c=\sqrt{\sigma/\rho g}$ is the capillary length, $\rho$ is the liquid density and $g$ is
the gravity acceleration. We assume that $f$ is periodic (period $2L$) in the $y$-direction
perpendicular to the direction of $u$. Following \cite{PomVan,Garoff}, the surface defects are
modeled by the spatial variation of the equilibrium value of the contact angle $\theta_{eq}(y,z)$
along the plate.

The contact line velocity with respect to the solid reads $v_n=\dot h+u$. Taking into account the
expression for the dynamic contact angle $\theta$ obtained under the condition (\ref{cond}),
\begin{equation}\label{cosdef}
\cos\theta=-\partial f/\partial x|_{x=0},
\end{equation}
one obtains from Eq. (\ref{cos1}) the following governing equation for $h$
\begin{eqnarray}\label{hL}
 \dot h(y)+u={\sigma\over\xi} \Biggl\{c[y, h(y)+ut]-\nonumber\\{1\over 2L}\sum_{n=-\infty}^\infty
 \sqrt{l_c^{-2}+\pi^2 n^2/ L^2}\int_{-L}^L
\textrm{d}y'h(y')\cos{\pi n(y-y')\over L}\Biggr\},
\end{eqnarray}
where $c(y,z)=\cos[\theta_{eq}(y,z)]$ is introduced for brevity.

Consider the contact line motion equation for an arbitrary defect pattern. It can be obtained from
Eq.~(\ref{hL}) when taking the limit $L\rightarrow\infty$:
\begin{eqnarray}
 \dot h(y)+u={\sigma\over\xi} \biggl\{c[y, h(y)+ut]-
 {1\over\pi}\int_0^\infty\textrm{d}p
 \sqrt{l_c^{-2}+p^2}\nonumber\\\int_{-\infty}^\infty
\textrm{d}y'h(y')\cos[p(y-y')]\biggr\}.\label{hh}
\end{eqnarray}
An equation in a very similar form have been already written \cite{JoRob}. However, the integration
order was inverted which resulted in a mathematically intractable expression. Eqs.~(\ref{hL},
\ref{hh}) reduce to those obtained in \cite{EuLet03} when $u=0$.

One can easily derive a simpler "long-wave limit" version of Eq.~(\ref{hh}) by expanding $h(y')$
around $h(y)$ in the Taylor series:\begin{equation}\label{hc}
 \dot h+u={\sigma\over\xi} \left[c(y, h+ut)-{h\over l_c}+{l_c\over 2}{\partial^2h\over \partial
 y^2}\right].
\end{equation}
Notice that this further simplification is fully consistent with the initial assumption
(\ref{cond}). This form of the governing equation clearly shows why one needs to account for the
gravity while considering the deformation of the initially straight contact line. When the gravity
influence tends to zero, $l_c$ increases and the gravity induced (second derivative) term that
describes the contact line deformation becomes dominating. In other words, the gravity influence on
the contact line deformation is important even in the large capillary length limit.

Consider now Eqs.~(\ref{hh},\ref{hc}) from the point of view of the theory of the interface
depinning \cite{Stanley,Fisher,LeDoussal}. They have the form (\ref{depin}), where the random term
$\eta$ is replaced by the random term $c$. In Eq.~(\ref{hc}), one recognises the well-studied (see
\cite{Krauth} and references therein) quenched Edwards-Wilkinson equation. However, the external
force $F$ is missing in both equations.

\section{External force}\label{sec4}

Generally, one cannot apply a force directly to the contact line to make it move. Probably the only
exception is a motion of a sessile drop on a solid with a wettability gradient. This special case
will not be considered here. In more common situation of homogeneous average wettability, the
contact line can be moved by either exerting a force at the fluid mass as a whole or moving the
solid with respect to the fluid similarly to the Wilhelmy balance experiments, using which this
force can be measured.

The additional force $F$ that acts on the Wilhelmy plate due to the presence of the contact line
(per unit plate width in $y$-direction) consists of two parts \cite{JoRob}: the contribution of the
interface tensions at the contact line and the "friction" force due to the energy
dissipation:\begin{equation}\label{F1} F={1\over 2L}\int_{-L}^L
\textrm{d}y\left\{\sigma_{LS}-\sigma_{GS}+\xi\left[\dot h(y)+u\right]\right\},
\end{equation}
where the surface tensions of the gas-solid ($\sigma_{GS}$) and liquid-solid ($\sigma_{LS}$)
interfaces are introduced. According to the Young formula,
$c(y,z)=(\sigma_{GS}-\sigma_{LS})/\sigma$. By using Eq.~(\ref{cos1}), one obtains the final
expression
\begin{equation}\label{F}
F=-{\sigma\over 2L}\int_{-L}^L \cos\theta(y)\,\textrm{d}y,
\end{equation}
which means that the force in $\sigma$ units at each time moment can be obtained by averaging the cosine of the
dynamic contact angle along the contact line. The expression (\ref{F}) have been used by several authors (see
e.g. \cite{Moul,Koplik}). This force can be measured directly by separating it out from viscous drag using
special experimental techniques \cite{Moul} and is presented as a counterpart of the external force $F$ in
Eq.~(\ref{depin}) for the case of contact line depinning.

One can see now clearly the difference between interface depinning and contact line depinning. For interface
depinning, the force $F$ enters directly into the governing equation~(\ref{depin}). It can be controlled,
imposed, and may take arbitrary value. For the contact line depinning, the external force does not enter
directly the governing equations~(\ref{hh},\ref{hc}). It is hardly possible to be controlled. However, it can be
measured when the velocity $u$ is imposed. It can be calculated using Eq.~(\ref{F}), according to which the
force (per unit plate width) is bounded by the surface tension value. This fact can explain the nonlinearity of
the $F(v)$ dependence observed in \cite{Moul} at large velocities. However, the model based on
Eqs.~(\ref{hh},\ref{hc}) cannot exhibit this saturation. Because of the conditions (\ref{cond}, \ref{cosdef}),
$|F|\ll \sigma$ was implicitly assumed during the derivation of Eqs.~(\ref{hh},\ref{hc}).

\section{Application to a periodic defect pattern}

\begin{figure}[tbh]
  \begin{center}
  \includegraphics[height=6cm]{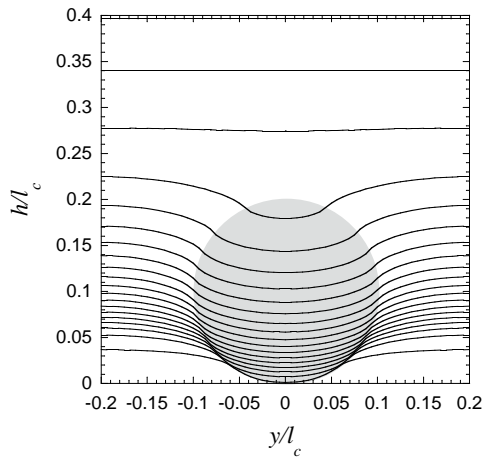}
  \end{center}
\caption{A unit cell for the periodic defect pattern (the area of a defect is shadowed) and
periodic (both in \emph{time} and space) solution of Eq.~(\ref{hL}). 20 snapshots of the contact
line with the equal time intervals $0.2\xi l_c/\sigma$ are shown for $v=0.1\sigma/\xi$. The chosen
parameters of the defect pattern are $2L=0.4l_c$, $r=0.1l_c$, $\theta_s=70^\circ$, and
$\theta_d=110^\circ$. The full picture of the contact line motion can be obtained by periodic
continuation of this image in both vertical and horizontal directions.} \label{Pos}
\end{figure}

We consider below a periodical both in the directions $y$ and $z$ pattern of round spots of the
radius $r$ shown in Fig.~\ref{Pos}.
Inside the spots, $\theta_{eq}=\theta_d$, the rest of the plate having $\theta_{eq}=\theta_s$.

Because of the nonlinearity in the $c$ term, Eq.~(\ref{hL}) seems to be complicated and difficult
to solve numerically. However, following considerations allow a quite efficient numerical algorithm
for its resolution to be proposed.

The function $c(y,z)$ can be made even with respect to $y$ by choosing properly the position of the
point $y=0$ with respect to the defect pattern. The function $h(y)$ is then even too and $\cos[\pi
n(y-y')/L]$ in Eq.~(\ref{hL}) factorises into $\cos(\pi ny/L)\cos(\pi y'/L)$. Both the integration
and the $n$-summation can be performed numerically with highly efficient Fast Fourier Transform
(FFT) algorithm \cite{NR}. The fourth-order Runge-Kutta method \cite{NR} is applied to solve the
differential (with respect to time) Eq.~(\ref{hL}). We are interested in its solutions periodic
both in $y$ and $t$. The time periodicity is sought to obtain time averaged values independent on
the initial position of the liquid surface. The time averages are denoted by the angle brackets,
e.g. the average force is
\begin{equation}\label{aver}
    \langle F\rangle=\frac{1}{P}\int_0^PF(t)\,\textrm{d}t,
\end{equation}
where $P=2L/|u|$ is the time period. The average contact line speed $v\equiv \langle v_n\rangle=u$.
The time-periodic behaviour appears after the contact line goes through several first rows of the
defects.

An example for such a double periodic solution is shown in Fig.~\ref{Pos}. The snapshots of the
contact line are "taken" with the equal time intervals, the contact line speed can be evaluated
from the density of the snapshots. One can see that when the contact line meets a line of defects,
its central portion remains stuck until the whole contact line slows down to let the liquid surface
accumulate its energy. During this stage, the difference between the dynamic and equilibrium
contact angles increases ("stick" stage). The slip stage follows, during which the contact line
accelerates. The difference between the average velocities in the stick and slip phases can be very
large near the pinning threshold, see the curve for $v=0.01\sigma/\xi$ in Fig.~\ref{F0_4} below,
where the most steep portion corresponds to the slip. This sequence of the accelerations and
decelerations of the whole contact line is a collective effect which characterises the contact line
motion in presence of defects.

The force (\ref{F}) can be calculated using Eq.~(\ref{cosdef}) for each of the $h(y)$ curves like
those in Fig.~\ref{Pos}. The $F(t)$ curves are presented in Fig.~\ref{F0_4} where $F$ is counted
from the value
\begin{equation}\label{fcb}
F_{CB}=\xi v-\sigma\cos\theta_{CB},
\end{equation}
where $\cos\theta_{CB}=\varepsilon^2\cos\theta_d+(1-\varepsilon^2)\cos\theta_s$ is the
Cassie-Baxter value of the static contact angle, and $\varepsilon^2=\pi (r/2L)^2$ is the defect
density. $F_{CB}$ corresponds to a force that would be induced by a homogeneous solid with the
equilibrium contact value equal to $\theta_{CB}$ which is simply a spatially averaged value of the
contact angle. It does not take into account the pinning on the defects. The difference $F-F_{CB}$
characterises the influence of the spatial fluctuations of $\theta_{eq}$ on the contact line
motion, i.e. the pinning strength.
\begin{figure}[tbh]
  \begin{center}
  \includegraphics[height=6cm]{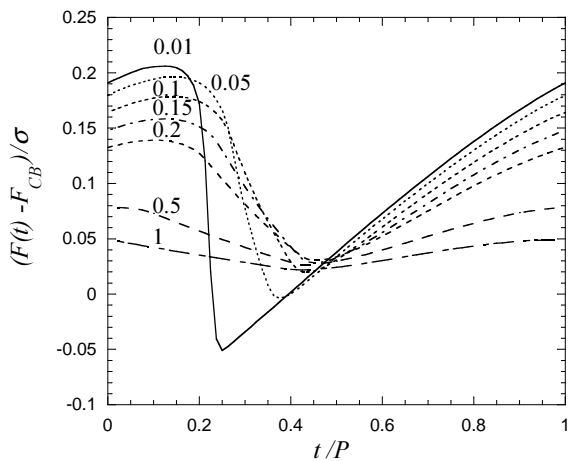}
  \end{center}
\caption{Periodic (with period $P$) variations of the force that acts at the Wilhelmy plate during its
downward motion. The parameter of the curves is $v$ in $\sigma/\xi$ units. We used the same parameters of the
defect pattern as for Fig.~\ref{Pos}.} \label{F0_4}
\end{figure}

\begin{figure}
  \begin{center}
  \small{(a)}\hspace*{7cm}\small{(b)}\\
  \includegraphics[height=6cm]{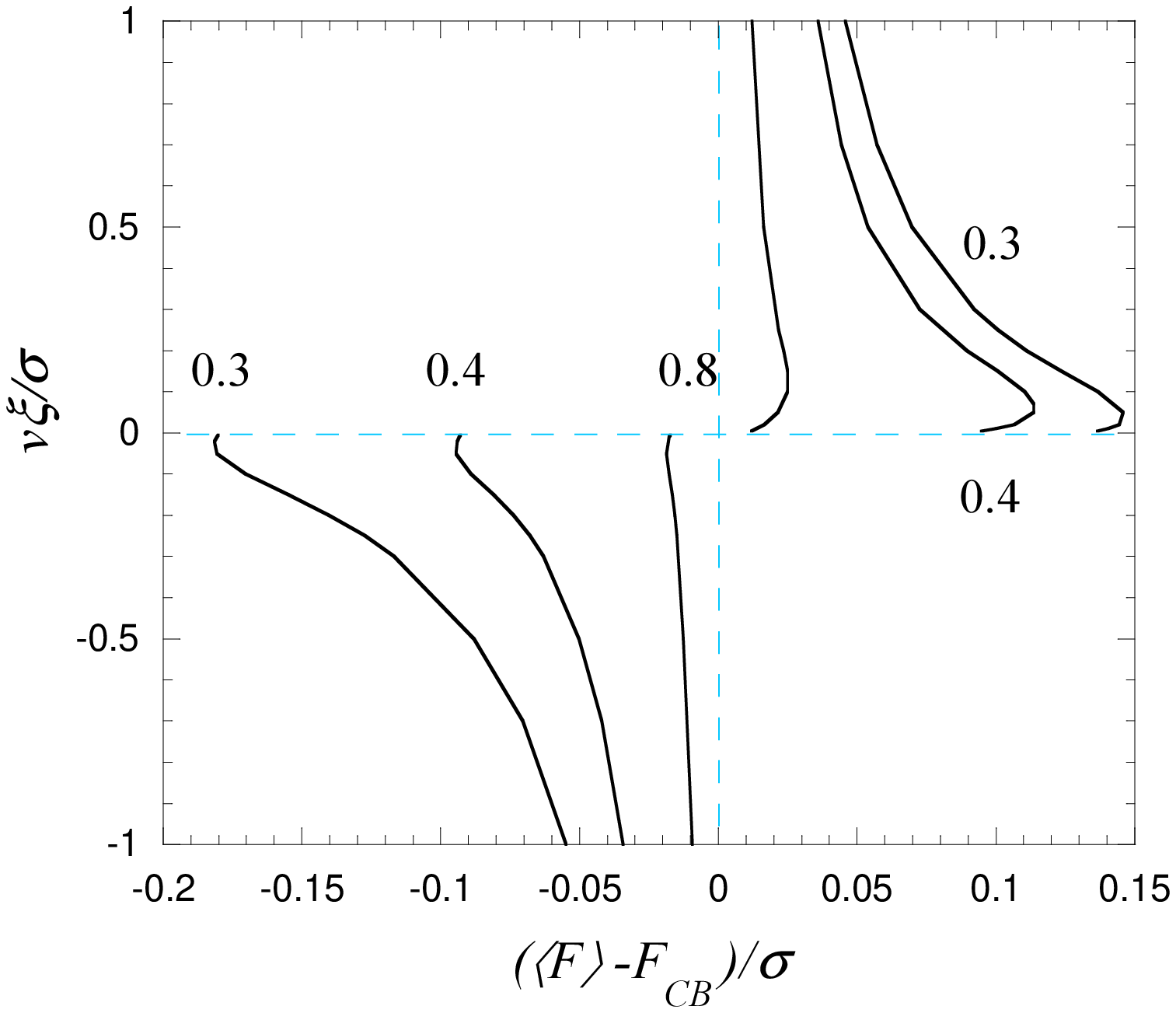}\includegraphics[height=5.8cm]{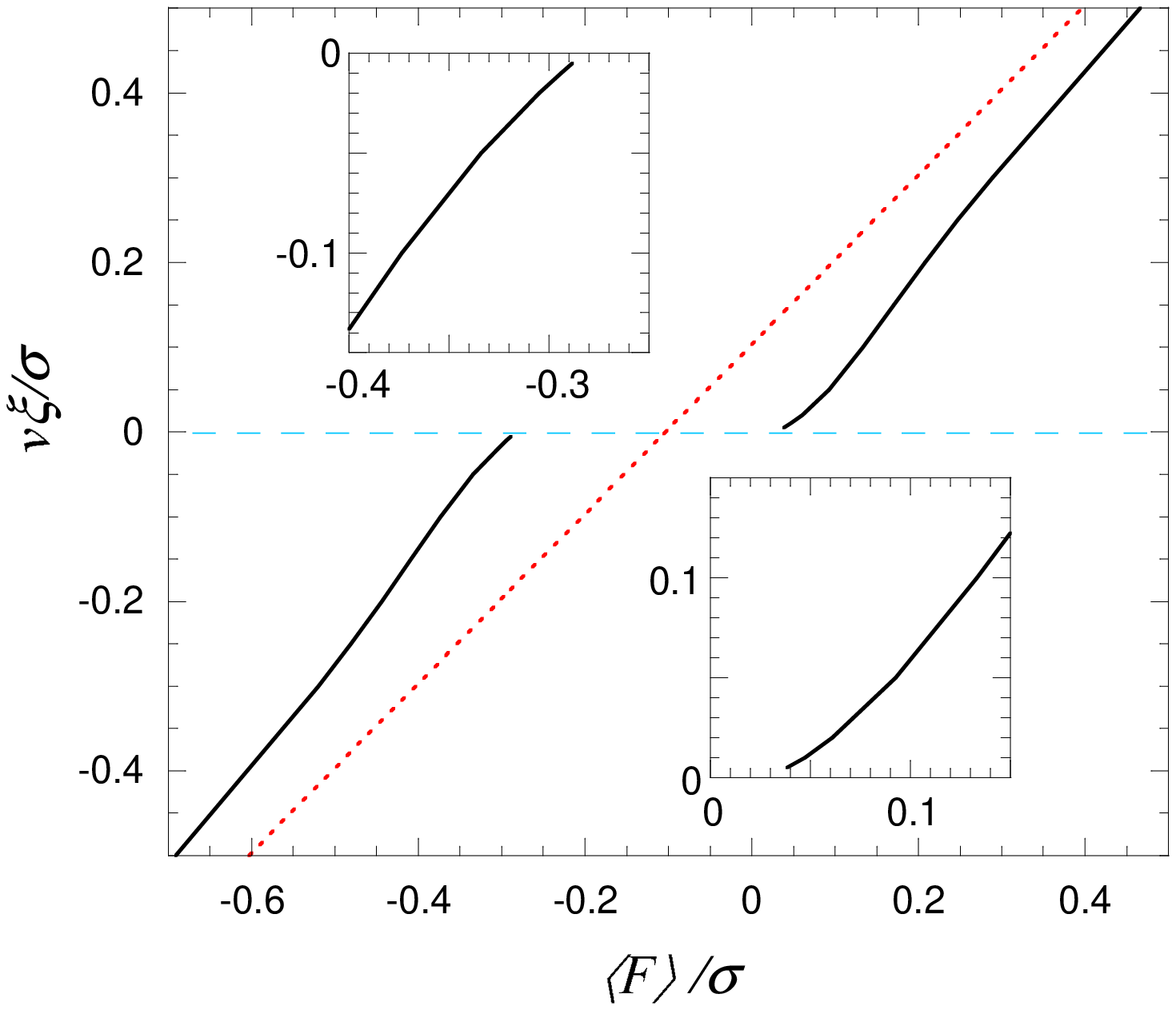}
  \end{center}
\caption{(a) $v(\langle F\rangle-F_{CB})$ curves calculated for different distances between defect
centres $2L$ (shown as a curve parameter in $l_c$ units). Both advancing ($v>0$) and receding
($v<0$) branches are presented. (b) $v(\langle F\rangle)$ curve for $2L=0.3l_c$. The $v(F_{CB})$
dependence is shown as a dotted line. The portions of the curves near $v=0$ are zoomed in the
inserts. Note that the abscissa is the averaged value of $\cos\theta$. Same parameters of the
defect pattern as for Fig.~\ref{Pos} are used.} \label{Fav}
\end{figure}

The dependence of $\langle F\rangle-F_{CB}$ on $v$ (inverted for compatibility with
Fig.~\ref{Fav}b) is shown in Fig.~\ref{Fav}a for different defect densities $\varepsilon^2$ that
correspond to different $L$ values. Both advancing ($v>0$) and receding ($v<0$) branches are
presented. The deviation of $\langle F\rangle$ from $F_{CB}$ increases with the increasing defect
density (decreasing distance between the defects) which is explained by the increasingly strong
pinning. By recalling that the average cosine of the contact angle is $\langle F\rangle/\sigma$,
one finds out that the cosines of the static advancing and receding contact angles (the values of
$\langle F\rangle/\sigma$ at $v\rightarrow \pm 0$) also drift away from the Cassie-Baxter value
with the increasing pinning.

One can notice some asymmetry of the force with respect to the direction of motion (advancing or
receding), which is visible in Fig.~\ref{Fav}a. In other words, $\cos\theta_{CB}\neq[\cos\theta(v)+
\cos\theta(-v)]/2$ in spite of the perfect symmetry of the pattern. This is explained by the
asymmetry of the problem geometry (Fig.~\ref{Wilh}) with respect to the motion direction.

One notices that the surface defects manifest itself much stronger at smaller velocities. It is
quite a general feature: at $|v|\geq\sigma/\xi$ the contact line does not "feel" the $\theta_{eq}$
fluctuations any more and the average cosine of the dynamic contact angle is defined by
$\cos\theta_{CB}-\xi v/\sigma$ for any defect pattern (until it attains the saturation regime at
$\cos\theta\approx\pm 1$, see sec.~\ref{sec4}).

While we study the pinning on the periodic patterns and the exponents proper to the random
behaviour cannot be recovered, it is however interesting to study the dependence of $\langle
F\rangle$ on $v$ and compare it to the behaviour defined by Eqs.~(\ref{vc},\ref{vf}). The inverse
functions $v(\langle F\rangle)$ are presented in Fig.~\ref{Fav}b . The $v(F_{CB})$ linear
dependence (inverted Eq.~(\ref{fcb})) is drawn for the sake of comparison.

Fig.~\ref{Fav}b can be compared to the experimental results \cite{Moul} obtained for the random defect pattern
(studies with the ordered patterns in this geometry are unknown to us). At small $|v|$, the sign of the
curvature is the same as in the experiment and corresponds to $\beta>1$ in Eq.~(\ref{vc}). This comparison
suggests that $\beta>1$ behaviour appears due to the collective pinning rather than due to the randomness.
Fig.~\ref{Fav}a shows that the calculations for the other defect densities exhibit the same curvature sign both
for the advancing and receding directions. The value of $F_c$ is defined by the static contact angle (advancing
or receding depending on the direction of motion). However, the linear increase of $\langle F\rangle(v)$ at
large $|v|$ similar to Eq.~(\ref{vf}) is simply a consequence of the approximation (\ref{cond}) discussed in the
previous section. In reality, the $v(\langle F\rangle)$ dependence is strongly nonlinear at large $|v|$ and
should have vertical asymptotes at $\langle F\rangle=\pm\sigma$.

The decreasing slope of the $v(F)$ curve at $F\rightarrow F_c$ (that appears due to the influence
of defects when $\beta>1$) can explain the extremely slow relaxation observed during the
coalescence of sessile drops \cite{JFM,Lang04}. It this case a very small force $F$ was imposed by
the surface tension. Since the effective dissipation coefficient was inferred from the $v(F)$ slope
value (inversely proportional to it), it appeared to be very large while the actual $\xi$ value
could be much smaller.

Previously, the $F(v)$ behaviour have been studied by Joanny and Robbins \cite{JoRob} for the 1D
case (they introduced an averaging along the $y$ axis) where $c$ was a function of $z$ (see
Fig.~\ref{Wilh}) only. The resulting from such a calculation contact line was thus always straight.
They considered several shapes for the $c(z)$ distribution and found that the $F(v)$ curvature sign
was different depending on the shape and periodicity of the $c(z)$ curve. For the square well shape
that would correspond to one studied here, they found the linear $F(v)$ dependence. In the present
study, $c$ varies also in $y$ direction, $c=c(y,z)$.

In this work we study an ordered defect pattern. Further studies will show if the universality in
the $v(F)$ law (\ref{vc}) exists for random defect patterns; it does not seem reasonable to us to
estimate the $\beta$ value at this stage.

\section{Conclusions}

It was demonstrated in this paper that the descriptions of the depinning of interface separating
two phases (e.g. for fluid invasion of porous media) and of the triple contact line, while similar
in many respects, has essential differences. The main of them is related to the external force that
can be controlled directly for the case of interface depinning and enters its equation of motion as
an additive term. An external force can hardly be imposed directly to the triple contact line and
thus does not enter its equation of motion. The experimentally measured force associated with the
contact line motion can be calculated and turns out to be essentially nonlinear in the contact line
velocity. At small velocities, the nonlinearity is due to the collective pinning at the surface
defects, while at large velocities the force per unit contact line length is bounded by the value
of the surface tension. Our theoretical results obtained for a periodical defect pattern suggest
that the experimentally observed \cite{Moul} nonlinearity of the force-velocity curve is a result
of the collective pinning on the defects rather than a consequence of their randomness.

This nonlinearity was obtained from the model with a constant dissipation coefficient $\xi$. Therefore, by
basing on the nonlinearity of this curve it is hardly possible to judge about the nonlinearity of the
microscopic mobility expression (i.e. dependence of the friction force that enters Eq.~(\ref{F1}) on velocity)
or, equivalently, on the dependence of $\xi$ on the contact line speed. Our results suggest that when the
external force is imposed (as for the elongated sessile drop returning to its final shape), the contact line
motion should be slowed down near the pinning and depinning thresholds.

The obtained profiles of the contact lines compare well to the recent experimental profiles
obtained with the periodic defect patterns for the sessile drop geometry \cite{Ferm}.

The equations of the contact line motion are derived. They can be applied to analyse the collective
effect of surface defects on the contact line motion for random defect patterns.

\ack The author would like to thank E.Rolley, S.Moulinet for useful discussions and D.Beysens for
numerous fruitful discussions and friendly support.

\end{document}